\documentclass[12pt]{iopart}
\input epsf

\newcommand{\gcc}{{\rm~g\,cm}^{-3}}
\newcommand{\msol}{\mbox{M}_\odot}

\newcommand{\ApJ}[1]{{\it Astrophys.~J.} {\bf #1}}

\begin{document}
\title[Dense astrophysical plasmas]{Dense astrophysical plasmas}

\author{G Chabrier\dag, F Douchin\dag\ and A Y Potekhin\ddag}
\address{\dag\ Ecole Normale Sup\'erieure de Lyon,
C.R.A.L.(UMR 5574 CNRS)\\
69364 Lyon Cedex 07, France }

\address{\ddag\ Ioffe Physico-Technical Institute, 194021 St Petersburg, Russia }

\begin{abstract}

We briefly examine the properties of dense plasmas characteristic of the atmospheres
of neutron stars and of the interior of massive white dwarfs. These astrophysical bodies are
natural laboratories to study
respectively the problem of
pressure ionization of hydrogen in a strong magnetic field and the crystallization of the
quantum one-component-plasma at finite temperature.
\end{abstract}

\maketitle

\begin {center} {prepared for ``Liquid state theory : from white dwarfs to colloids''\\
An international conference on the occasion\\ of Prof. Jean-Pierre 
Hansen's 60th birthday\\
Les Houches April 1-5, 2002}
\end {center}


\section{Introduction}

The quest for an accurate description of the thermodynamics properties of dense plasmas has represented a thriving domain
of research since the seminal review by Baus {and} Hansen (1980).
The interiors
 of many astrophysical bodies are characterized by strongly 
correlated ionic and electronic plasmas, with respective classical and quantum coupling parameters
$\Gamma_{\rm i}=\beta(Z_{\rm i}e)^2/a_{\rm i}$ and $r_s=a_{\rm i}/(a_0 Z_{\rm i}^{1/3})$ 
varying over several decades. Here, $a_0=\hbar^2/(m_e e^2)$ denotes the electronic Bohr radius,
 $a_{\rm i}=(3N_{\rm i}/4\pi V)^{1/3}$ is the mean inter-ionic distance,
 and $\beta\equiv 1/( k_{\rm B}T)$.
The correct description of the thermodynamic
properties of these astrophysical bodies, which determine their mechanical and thermal properties, thus requires the knowledge of the equation of state (EOS) of such plasmas.
In this short review, we focus on the case of neutron stars (NS) and massive white
dwarfs (WD), which exhibit two particularly interesting problems in the
 statistical physics of dense matter.

\section{Ionization equilibrium 
of a hydrogen plasma in strong magnetic fields}

Most of neutron stars are characterized by magnetic fields $B\sim10^{11}$--$10^{13}$~G, whereas some of them (so-called magnetars)
are thought to have $B\sim10^{14}$--$10^{15}$~G.
Although huge by the terrestrial standards, 
the magnetic energy of a NS,
$\sim R^3 B^2/6$, represents only
a tiny fraction of its 
gravitational binding energy $E_G\sim GM^2/R^2$.

The photospheric properties of a NS are characterized by temperatures 
$T\simeq 10^5$--$10^7$ K (depending on the age and mass of the star)
and densities
$\rho\simeq 10^{-2}$--$10^4\gcc$ (depending on $T$ and $B$). 
The cooling rates of these
stars are entirely determined by the relationship between the photospheric and the interior temperature profiles. 
The emitted spectra of these stars
can be strongly affected by 
the presence of bound species in the photosphere.
Since the atmosphere can be composed essentially of hydrogen accreted from either the
interstellar medium, the supernova remnant or a close companion, the determination of the temperature profiles and spectra
thus requires an accurate description of hydrogen ionization 
in a strong magnetic field.

The quantum-mechanical properties of protons, free electrons
and bound species (hydrogen atoms and molecules) are strongly modified by
the field, which thereby affects the thermodynamic properties of the plasma.
The properties of matter in a magnetic field under NS conditions
have been reviewed
recently by Ventura and Potekhin (2001) and Lai (2001).
We refer the reader to these reviews for the detailed descriptions of these properties, and only the most recent results will be outlined in the present paper.

The transverse motion of electrons in a magnetic field
is quantized into Landau levels. The energy of the $n$th
Landau level of the electron (without the rest energy) is
$m_e c^2(\sqrt{1+2bn}-1)$,
which becomes $\hbar\omega_c n$ in the non-relativistic limit,
where
\begin{equation}
\hbar \omega_c= \hbar {eB\over m_e c} = 11.577 B_{12} \,{\rm keV},
\label{eqn_cyclo}
\end{equation}
is the electron cyclotron energy,
\begin{equation}
b=\hbar\omega_c/m_e c^2=B_{12}/44.14
\end{equation}
is the field strength in the relativistic units,
and $B_{12}=B/(10^{12}\,{\rm G})$ is a typical 
magnetic-field scale for NS conditions.

The atomic unit for the magnetic-field strength is set by $\hbar \omega_c=e^2/a_0$,
i.e. $B_0=(m_ec/\hbar e)\times(e^2/a_0)=2.35\times 10^9$ G.
It is convenient to define a dimensionless magnetic-field strength
\begin{equation}
\gamma= {B/ B_0}=b/\alpha_f^2\,,
\end{equation}
where $\alpha_f$ is the fine structure constant.

For $\gamma\gg 1$, as encountered in NS conditions, the electron cyclotron energy is much larger than the
typical Coulomb energy, so that the properties of interacting particles, protons, atoms, molecules, are strongly affected by the field.
The ground-state atomic and molecular binding energies increase with $B$
as $\sim\ln^2\gamma$.
The atom in a strong magnetic field is compressed in the transverse directions to the radius $\sim a_{\rm m}$, where
\begin{equation}
a_{\rm m}= ({\hbar c / e B})^{1/2}= \gamma^{-1/2} a_0 = 2.56\times 10^{-10} B_{12}^{-1/2} \,{\rm cm}
\label{eqn_cyclorad}
\end{equation}
is the quantum magnetic length, which
becomes the natural length unit in the plasma instead of $a_0$.

The thermal motion of atoms causes the Stark effect due to
the electric field induced in the comoving frame of reference.
At $\gamma\gg1$, this effect strongly modifies the atomic 
properties: the atom becomes asymmetric, and its binding energy
and oscillator strengths
depend on the velocity (Potekhin 1994).
Two classes of the atomic states arise: so-called centered and
decentered states; for the latter ones the electron-proton 
separation is large and the binding energy relatively small
 (Vincke \etal 1992; Potekhin 1994).

The formation of molecules is also strongly modified in a strong magnetic field.
Because of the alignment of the electron spins antiparallel to the field, two atoms in their ground state
($m=0$) do not bind together, because of the Pauli exclusion principle. One of the two H atoms has to be excited in the $m=-1$ state to form the ground state of the
H$_2$ molecule, which then forms by covalent bonding (e.g., Lai, 2001). 

As long as $T\ll \hbar\omega_c/k_{\rm B} = 1.343\times10^8\,B_{12}$~K and
$\rho \ll \rho_B \approx 7.1\times10^3\,B_{12}^{3/2}\gcc$,
the electron cyclotron energy $\hbar \omega_c$ exceeds both the thermal energy
$k_{\rm B}T$ and the electron Fermi energy $k_{\rm B} T_{\rm F}$, 
so that the electrons are mostly in the Landau ground state, -- i.e.,
the field is {\it strongly quantizing}. 
In this case, typical for the NS photospheres,
the electron spins are aligned antiparallel to the field. 

Proton motion is also quantized by the magnetic field,
but the corresponding cyclotron energy is smaller,
$\hbar\omega_{cp}= \hbar\omega_c m_e/m_p$.

\begin{figure}
\begin{center}
\epsfxsize=145mm
\epsfbox{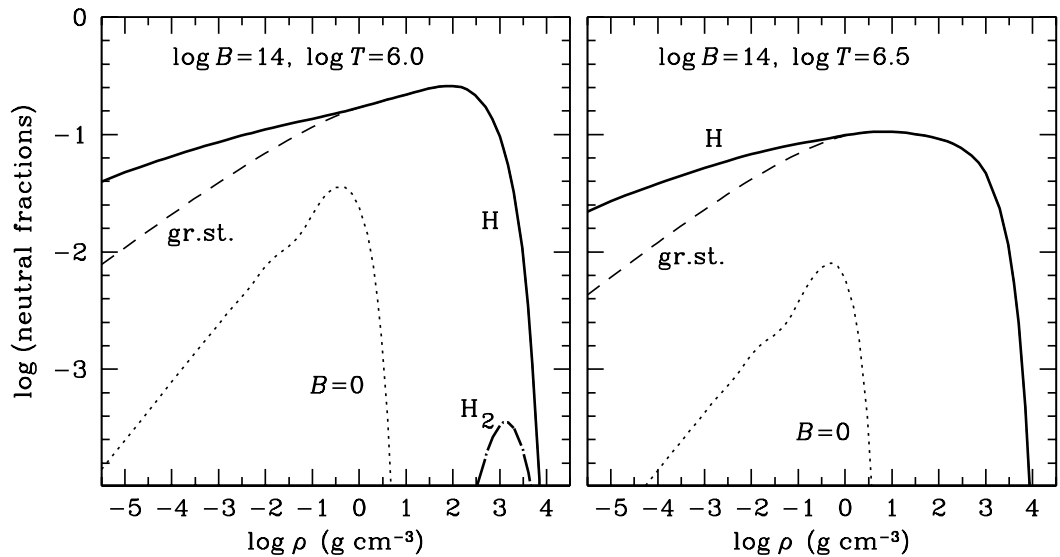}
\end{center}
\caption{Ionization isotherms at $B=10^{14}$~G and two values of $T$.
Solid lines: total fraction of atoms $f_{\rm H}=n_{\rm H}/n_0$;
dash-lines: fractions of ground-state atoms;
dot-dashed lines: molecular fraction $f_{{\rm H}_2}=2n_{{\rm H}_2}/n_0$,
where $n_0$ is the total number density of protons (free and bound).
Dotted lines: zero-field case}
\end{figure}

Quantum-mechanical calculations of the
binding energies and wave functions
of hydrogen atoms in {\em any\/} states of motion
in the strong magnetic
fields have been carried out only recently (Vincke \etal 1992; Potekhin, 1994).
Using these results, Potekhin, Chabrier and Shibanov (1999) derived
a model which describes the 
thermodynamics
of an interacting (H$_2$, H, H$^+$, e$^-$) plasma in a strong magnetic field.
This model is based on the framework of the free energy model
developed by Saumon {and} Chabrier (1991, 1992) for pressure ionization of hydrogen at $B=0$, but generalizes it to the strong-$B$ case,
taking into account the quantum-mechanical effects 
caused by the thermal motion of atoms across the magnetic field.

Potekhin \etal\ (1999) calculated the ionization equilibrium and EOS
at $7\times10^{11}{\rm~G} \leq B \leq 3\times10^{13}$~G.
In this paper, we extend these calculations  
 up to $B=10^{14}$~G,
 typical for the magnetars. We calculate the number densities
  of atoms ($n_{\rm H}$) and molecules ($n_{{\rm H}_2}$)
from the equations
\begin{equation}
\fl
    n_{\rm H} =  n_e^2  \,                                     
   { \lambda_p\lambda_e (2\pi a_{\rm m}^2)^2
     \over \lambda_{\rm H}^3 } 
    \, \left[1-e^{-\beta\hbar\omega_{cp}}\right] \, 
    Z_w {\rm e}^{\Lambda},
\quad 
    n_{{\rm H}_2} = n_{\rm H}^2 (\lambda_{\rm H} \sqrt{2})^3 
    Z_{w2} / Z_w^2, 
\end{equation}
where $n_e$ is the electron number density,
$\lambda_j =(2\pi \beta\hbar^2/m_j)^{1/2}$
is the thermal wavelength of the particle $j$
($j=e$, $p$, H), $Z_w$ and $Z_{w2}$ are the internal partition 
functions for H and H$_2$, respectively,
and $\Lambda$ is a correction factor, which takes into account
electron degeneracy and filling of the excited Landau levels.
The formulae for $Z_w$, $Z_{w2}$ and $\Lambda$ 
are given in Potekhin \etal (1999).

Figure 1 displays the logarithm of the fraction of H and H$_2$ for a magnetic field
$B=10^{14}$ G, for two isotherms. The solid line represents the total fraction
of atoms $f_{\rm H}=n_{\rm H}/n_0$ ($n_0=n_{\rm H}+n_p+2n_{{\rm H}_2})$
in all quantum states, whereas
the dashed line shows the fraction of atoms in the ground
state. The dotted line displays the zero-field result. 
As seen in this figure, the strong magnetic field favours atomic and molecular recombination.
Since the binding energies of atoms and molecules increase 
and $T_{\rm F}$ decreases in the strongly quantizing magnetic field,
 pressure ionization
occurs at much larger densities than for the field-free case.

\begin{figure}
\begin{center}
\epsfxsize=145mm
\epsfbox{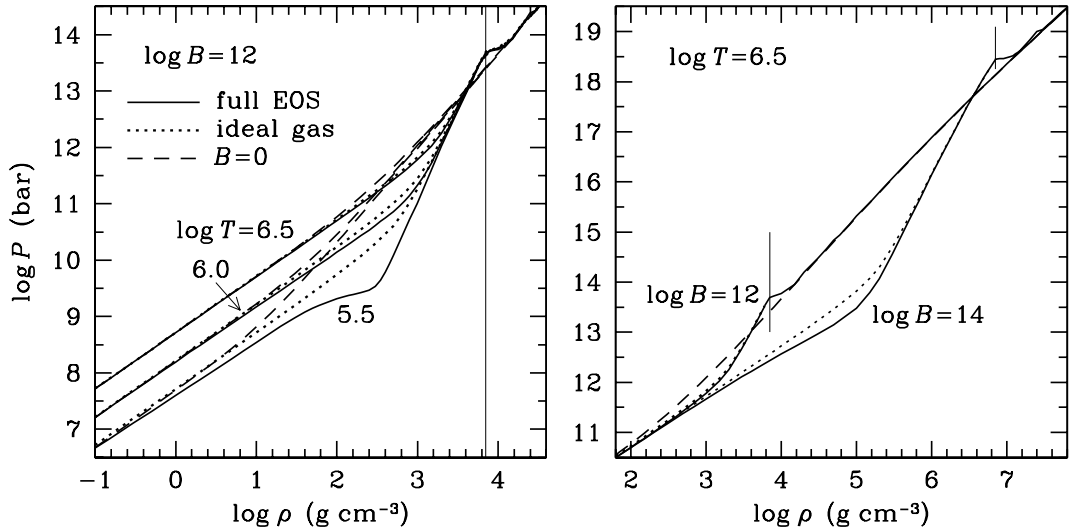}
\end{center}
\caption{EOS of partially ionized atomic hydrogen 
in the strong magnetic field (solid lines)
compared with the EOS of a fully ionized ideal
electron-proton plasma (dotted lines)
and the EOS of a partially ionized hydrogen at $B=0$
(dashed lines). Left panel: $B=10^{12}$~G,
$\log T\mbox{[K]} = 6.5, 6.0$, and 5.5.
Right panel: $\log T\mbox{[K]} = 6.5$,  $B=10^{12}$~G
and $10^{14}$~G.
The vertical lines correspond to the density above which
excited Landau levels become populated}
\end{figure}

Figure 2 displays the resulting EOS. The left panel displays
three isotherms at $B=10^{12}$~G, typical of ordinary NS. 
On the right panel, we compare the isotherm $T=10^{6.5}$~K
for $B=10^{12}$~G and for the superstrong field $B=10^{14}$~G.
As expected from the previous discussion, 
the EOS in a strong magnetic
field differs significantly from the field-free case in the region of partial ionization.
At very high-density, excited Landau levels become populated, due to the increase of the Fermi energy, and the zero-field case is recovered.

\section{Crystallization of white dwarf cores}

Massive white dwarfs ($1.2\,\msol \le M \le 1.4\, \msol=M_{\rm Ch}$, where $M_{\rm Ch}$
denotes the Chandraskhar mass) are C$^{6+}$/O$^{8+}$ plasmas
with central density and temperatures 
$\rho_{\rm c} \approx$10$^8$--10$^9\gcc$, $T_{\rm c}\approx 10^6$ K. Under these conditions, the ion zero-point energy $E_0 \propto \hbar \Omega_P $, where
$\Omega_P=(3Z_{\rm i}^2e^2/M_{\rm i}a_{\rm i}^3)^{1/2}$ is the ion plasma frequency, exceeds the classical
thermal energy $k_{\rm B}T$ (Chabrier \etal 1992), i.e. $\eta =\hbar \Omega_P/k_{\rm B}T \gg 1$. Collective
diffraction effects thus modify the classical Coulomb interaction. The melting values of the coupling parameters $\Gamma_{\rm m}$  for the classical OCP and $R_{S,{\rm m}}$, where $R_S=a_{\rm i}/(\hbar^2/M_{\rm i} Z_{\rm i}^2e^2)$ is the ion
quantum plasma parameter, for the quantum jellium model at zero-temperature, have been firmly established: $\Gamma_{\rm m}=175$ (Potekhin {and} Chabrier 2000), and $R_{S,{\rm m}}=160$  for bosons, $R_{S,{\rm m}}=100$ for fermions (Ceperley {and} Alder 1980). The melting curve at {\it finite temperature}, however, 
i.e., at $\eta \ne 0$ and $T\ne 0$, still remains poorly determined.

A simplified determination of this curve is based on
 the Lindeman critical parameter
interpolation between the zero-temperature and the classical melting values (Mochkovitch {and} Hansen 1979; Chabrier 1993). These calculations,
however, are based on a harmonic description of the phonon mode spectrum and thus do not include
non-harmonic effects. More recently, Jones {and} Ceperley (1996) 
performed Path Integral Monte Carlo (PIMC) simulations to try to determine the
phase diagram more correctly. They found a maximum melting temperature almost a factor of  2 larger than the one determined by Chabrier (1993).

In this paper, we present preliminary results based on similar PIMC calculations
at finite temperature. 
The system under consideration
consists 
 of $N$ identical,  but {\it distinguishable } particles (Boltzmannions)
with a mass $M$ and
charge $Ze$ in a volume $V$ at temperature $T$.

 The partition function of the quantum system is given by
the trace of the $N$-body density matrix $\rho_N(\vec R,\vec R^\prime;
\beta) = \langle\vec R| e^{-\beta H_N}| \vec R^\prime \rangle$.
Here $\vec R$ denotes the $3N$ coordinates $\{ \vec r_i\}_{i=1,...,N}$.
Using the Trotter formula, this  partition function 
can be rewritten exactly in terms of the density matrix 
$\rho_N(\vec R,\vec R^\prime;\tau) $ 
with 
$\tau=\beta/P$ (Ceperley 1995) as
\begin{equation}
Q_{N,V,T}(\beta)=\int \Pi_{\alpha=1}^P d\vec R_{\alpha} \rho_N
(\vec R_{\alpha},\vec R_{\alpha +1};\tau)
\end{equation}
In the $P\rightarrow \infty$ limit, the exact density matrix
\begin{equation}
e^{-\beta H_N}={\rm lim}_{P\rightarrow \infty} [e^{-{\beta \over P}K_N} \times e^{-{\beta \over P}V_N}]^P
\end{equation}
 is recovered by using the approximate expression
\begin{equation}
\rho_N(\vec R_{\alpha},\vec R_{\beta};\tau)=\rho_N^0(\vec R_{\alpha},
\vec R_{\beta};\tau)\, 
\exp\left[ -{\tau \over 2} \left\{ V^c(\vec R_{\alpha})+
V^c(\vec R_{\beta})\right\} \right]\,,
\end{equation}
where $K_N$ and $V_N$ denote the kinetic and potential parts of the Hamiltonian, respectively, 
$\rho_N^0(\vec R_{\alpha},\vec R_{\beta};\tau)=
({4\pi \tau})^{-3N/2} \exp[-(\vec R_{\alpha}-\vec R_{\beta})^2] /(4\tau)$
 is the density matrix of free particles, 
 and $V^c$ is the classical potential energy of the system.

In terms of the path integral formalism, the particle is defined by its trajectory in ``imaginary time'' $P\tau$, through ``polymers'' composed of $P$ monomers connected by ``strings'' of stiffness
$M_{\rm i}/\hbar\tau$.

We have performed simulations along three isochores $R_S=1200,350,200$ and five isotherms
on each isochore around the estimated melting temperature. In order to estimate the finite-size
effects, we used 16, 54, 128 and 250 particles. For each simulation we take $\eta/P=$0.05--0.1.
We then parametrized the fluid and solid internal
energies, including the finite-size corrections. The melting temperature was estimated both from
free energy comparison and from a dynamical criterion, namely the mean square displacement
in unit of the nearest-neighbour distance $(\langle \delta r^2 \rangle / d^2)^{1/2}$. This value is finite in a solid
and diverges in a fluid.

\begin{figure}
\begin{center}
\epsfxsize=95mm
\epsfbox{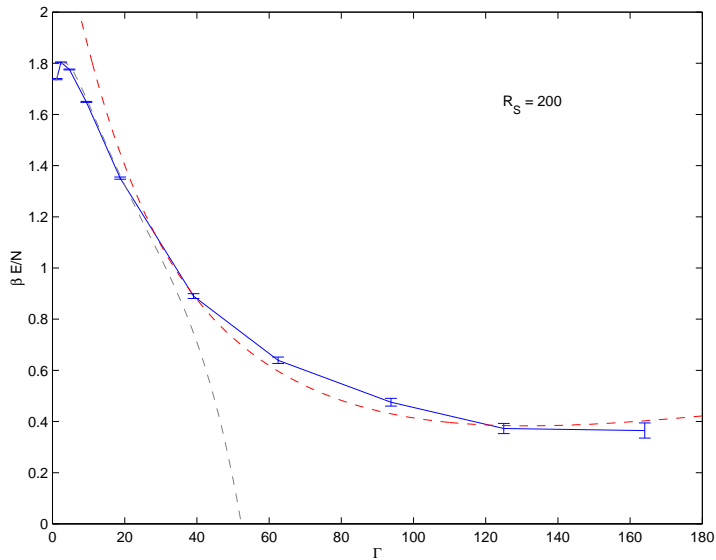}
\end{center}
\caption{Internal energy of a quantum OCP at $R_S=200$.
Solid lines: PIMC calculations;
left dash-line: Hansen-Vieillefosse (1975) $\hbar^4$-expansion;
right dash line: analytical fit of the energy at large $\Gamma$ (Douchin {and} Chabrier 2002).
}
\end{figure}

Figure 3 displays the internal energy of the quantum fluid for $R_S=200$. The solid line is the
result of the PIMC simulation, with $N=54$ particles. The dash line displays the Hansen-Vieillefosse (1975)
$\hbar^4$-expansion, which diverges when the thermal wavelength $\lambda_{\rm i}$
is of the order of the inter-ionic
distance $a_{\rm i}$. At large $\Gamma$, i.e. large $\eta$ for
a given $R_S$,
the energy tends towards the zero-temperature limit (Chabrier, 1993). The dash-line
displays an analytical fit of the PIMC energy at large $\Gamma$ (Douchin {and} Chabrier, 2002).

Our preliminary results yield melting temperatures lower than the ones obtained by Jones {and} Ceperley (1996), about a factor of 2 for the maximum melting temperature. This difference stems from
a better correction of finite-size effects, which are known to stabilize the solid. First of all, we explored more in detail the N-dependence of our PIMC
simulations by conducting calculations with a larger number of particles (up to
N=256). 
Second of all, we found out that the dependence on $\Gamma $ and $R_S$ of the
finite-size effects for the liquid phase is more complex than the simple
one used by Jones and Ceperley (1996). This is particularly important for the
extrapolated values of the energy around the turning point.
The obtained maximum melting temperature and corresponding density are (Douchin {and} Chabrier, 2002)
\begin{equation}
 T_{\rm max}=8810\,A\,Z^4 \,\,{\rm K};
\quad
 \rho_{\rm m}=1280\,A^4\,Z^6 \,\gcc,\quad\mbox{i.e.,} \,\,R_S\approx 235\,.
 \end{equation}

These calculations show that, although anharmonic effects are non-negligible both in the
solid and in the liquid phase, they are comparable in both phases and almost cancel out, so that the melting curve lies
close to the one estimated from the harmonic spectrum.

\section{Conclusion}

In this short review, we have considered two different problems related to dense plasma
physics as encountered under specific astrophysical conditions. Because of the presence of a strong magnetic field, the quantum internal levels of atoms and molecules are quantized in Landau
orbits, and the field raises the binding energy of these species, favoring recombination
over dissociation and ionization compared with the field-free case. This modifies the ionization equilibrium and
EOS of the dense plasma.

We also considered the crystallization of a quantum fluid of Boltzmannions at finite temperature
with PIMC simulations, taking into account the correction due to finite-size effects. We found that the
crystallization diagram lies close to the one based on an interpolation of the Lindeman criterium between the classical and the zero-temperature limits. 
These preliminary results need to be confirmed by further calculations. Such work is under progress.

These two examples, and many other not mentioned in the present review,
 stress the need for detailed calculations of the properties of dense plasmas
under extreme conditions for astrophysical applications.

\ackn
The work of A.P.\ was supported in part 
by RFBR Grants 02-02-17668 and 00-07-90183.

\section*{References}
\begin{harvard}

\item[] Baus M {and} Hansen J P 1980 {\it Phys.\ Rep.} {\bf 59} 1

\item[] Ceperley D M 1995 \RMP {\bf 67} 279

\item[] Ceperley D M {and} Alder B J 1980 \PRL {\bf 45} 566

\item[] Chabrier G 1993 \ApJ{414} 695

\item[] Chabrier G, Ashcroft N W {and} DeWitt H E 1992 {\it Nature} {\bf 360} 48

\item[] Douchin F {and} Chabrier G 2002 (in preparation)

\item[] Hansen J P {and} Vieillefosse P 1975 \PL {\bf 53A} 187

\item[] Jones M D {and} Ceperley D M 1996 \PRL  {\bf 76} 4572

\item[] Lai D 2001 \RMP {\bf 73} 729



\item[] Mochkovitch R {and} Hansen J P 1979 \PL {\bf A73}, 35

\item[] Potekhin A Y 1994 \jpb {\bf 27} 1073

\item[] Potekhin A Y {and} Chabrier G 2000 \PR{\it E} {\bf 62} 8554
 
\item[]  Potekhin A Y, Chabrier G {and} Shibanov Yu A 1999 \PR{\it E} {\bf 60} 2193

\item[] Saumon D {and} Chabrier G 1991 \PR {\it A} {\bf 44}, 5122

\item[] Saumon D {and} Chabrier G 1992 \PR {\it A} {\bf 46}, 2084

\item[] Ventura J {and} Potekhin A Y 2001 
{\it The Neutron Star -- Black Hole Connection, NATO Science Ser.\ C} {\bf 567},
C Kouveliotou, J Ventura {and} E P J van den Heuvel
(Dordrecht: Kluwer) p 393

\item[] Vincke M, Le Dourneuf M and Baye D 1992 \jpb{\bf 25} 2787

\end{harvard}

\end{document}